\documentclass[prb,twocolumn,showpacs,superscriptaddress]{revtex4-2}
\usepackage{graphicx}
\usepackage{mathrsfs}
\usepackage{dcolumn}
\usepackage{bm}
\usepackage{amsmath}
\usepackage{amsfonts}
\usepackage{color}

\begin{document}

\title{A numerical method for designing topological superconductivity induced by s-wave pairing}

\author{Jingnan Hu}
\affiliation{Wuhan National High Magnetic Field Center $\&$ School of Physics, Huazhong University of Science and Technology, Wuhan 430074, China}
\author{Aiyun Luo}
\affiliation{Wuhan National High Magnetic Field Center $\&$ School of Physics, Huazhong University of Science and Technology, Wuhan 430074, China}
\author{Zhijun Wang}
\affiliation{Beijing National Laboratory for Condensed Matter Physics, and
Institute of Physics, Chinese Academy of Sciences, Beijing 100190, China}
\affiliation{University of Chinese Academy of Sciences, Beijing 100049, China}
\author{Jingyu Zou}
\affiliation{Wuhan National High Magnetic Field Center $\&$ School of Physics, Huazhong University of Science and Technology, Wuhan 430074, China}
\author{Quansheng Wu}
\email[e-mail address: ]{quansheng.wu@iphy.ac.cn}
\affiliation{Beijing National Laboratory for Condensed Matter Physics, and
Institute of Physics, Chinese Academy of Sciences, Beijing 100190, China}
\affiliation{University of Chinese Academy of Sciences, Beijing 100049, China}
\author{Gang Xu}
\email[e-mail address: ]{gangxu@hust.edu.cn}
\affiliation{Wuhan National High Magnetic Field Center $\&$ School of Physics, Huazhong University of Science and Technology, Wuhan 430074, China}
\affiliation{Institute for Quantum Science and Engineering, Huazhong University of Science and Technology, Wuhan, 430074, China}
\affiliation{Wuhan Institute of Quantum Technology, Wuhan, 430074, China}

\begin{abstract}
Topological superconductors have garnered significant attention due to their potential for realizing topological quantum computation. However, a universal computational tool based on first-principles calculations for predicting topological superconductivity has not yet been fully developed, posing substantial challenges in identifying topological superconducting materials. In this paper, we present a numerical method to characterize the superconducting spectrum and topological invariants of two-dimensional (2D) slab systems using first-principles calculations, implemented in the open-source software WannierTools. To more accurately model the superconducting proximity effect, we integrate an SC pairing decay module into the program. Our approach can be applied to classical superconductor–topological insulator (SC-TI) heterostructures, SC-semiconductor heterostructures, and intrinsic topological superconductors. The program’s validity is demonstrated using the topological crystal insulator SnTe, the Rashba semiconductor InSb, and the superconductor NbSe$_2$ as examples. We anticipate that this tool will accelerate the discovery of topological superconductor candidates.
\end{abstract}

\maketitle

\section{Introduction}

The concept of topological superconductors was first introduced in 2000 through theoretical models by Read and Green in two dimensions (2D)\cite{read2000paired} and by Kitaev in one dimension (1D)\cite{kitaev2001unpaired}. These models describe spinless, time-reversal-breaking p-wave superconducting states, which support non-Abelian Majorana zero modes. In the 2D case, these modes are localized in the vortex cores, while in the 1D case, they appear at the edges of the system. The existence of Majorana zero modes in these topologically protected states is considered a key feature for potential applications in topological quantum computation.

Interest in topological superconductors (TSCs) has grown significantly in recent years, driven largely by their connection to Majorana fermions—exotic particles that are their own antiparticles \cite{majorana1937teoria, alicea2012new, beenakker2013search, sato2016majorana}. These Majorana zero modes are particularly attractive for topological quantum computation due to their potential as stable qubits, leveraging their unique non-Abelian statistics to enable fault-tolerant quantum information processing \cite{bravyi2002fermionic, fan2006newton}. More recently, chiral Majorana states (MSs) have also been explored for similar applications, further expanding the possibilities for quantum computation \cite{lian2018topological}. As a result, TSCs have garnered significant attention from the academic community \cite{ivanov2001non, Sau2011controlling, zhang2013time, yang2014dirac, xu2014topological, kawakami2015evolution, zhang2019helical, zhang2021intrinsic, zou2021new, giwa2021fermi, margalit2022chiral}. Traditionally, TSCs have been associated with p-wave superconductors, which are predicted to support Majorana zero modes \cite{Maeno1994Superconductivity, nelson2004odd, Hor2010Superconductivity, jiao2020chiral, li2021boundary}. However, the intrinsic occurrence of p-wave superconductivity is rare, and the direct observation of Majorana states in such systems has proven challenging. Over the past decade, numerous strategies have been proposed to induce TSCs at the surfaces or interfaces of topological materials, including semiconductors with Rashba spin-orbit coupling (Rashba-SOC) \cite{sato2009non, lutchyn2010majorana, oreg2010helical, sau2010generic, alicea2010majorana, stanescu2011majorana}, magnetic atom chains \cite{nadj2013proposal, klinovaja2013topological, kim2014helical, li2014topological, glazov2014exciton}, and obstructed atomic insulators \cite{hu2024chiral, sheng2024majorana}, all of which utilize the proximity effect with s-wave superconductors. Additionally, an alternative approach focuses on the realization of TSCs within superconducting topological metals, which combine topologically nontrivial electronic structures at the Fermi level with superconductivity in a single material. Examples of such systems include iron-based superconductors \cite{xu2016topological, zhang2018observation, wang2018lingyuan, liu2018robust, liu2020new, kong2021majorana, li2022ordered}. This growing body of research highlights the importance of understanding the mechanisms and conditions under which TSCs can be realized, offering promising avenues for their application in next-generation quantum technologies.

\begin{figure}[t]
    \centering
    \includegraphics[width=0.5\textwidth]{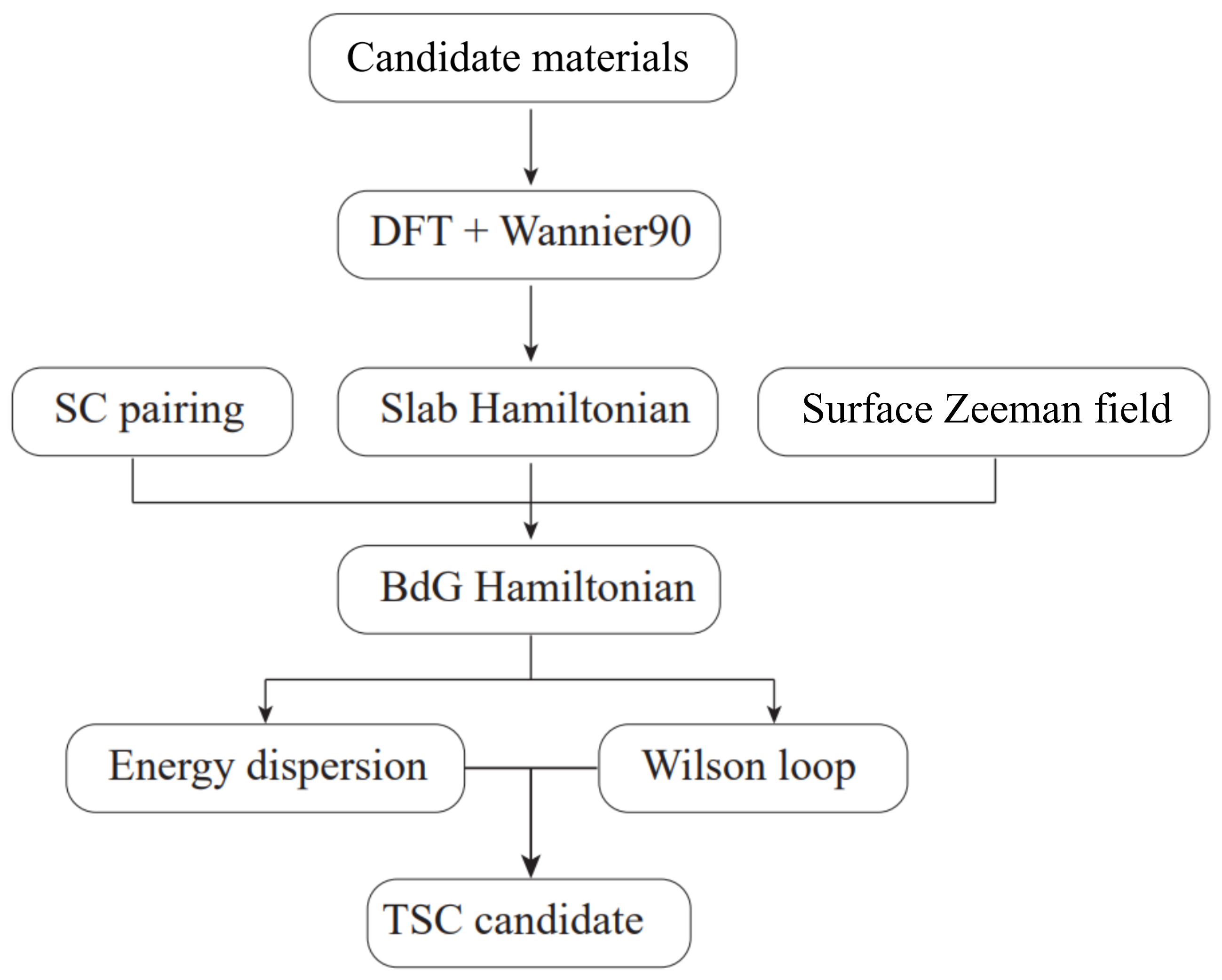}
    \caption{Flow chart illustrating the numerical method developed to identify candidate materials for topological superconductors (TSCs). The chart outlines the sequential steps involved, starting from first-principles calculations to the construction of Wannier functions, the introduction of superconducting pairing and Zeeman splitting, and culminating in the analysis of the superconducting spectrum and topological invariants. }
    \label{program flow chart}
\end{figure}

Despite significant theoretical progress, experimental confirmation of topological superconductors (TSCs) remains relatively limited. A primary challenge lies in the fact that theoretical predictions are often based on simplified, effective models, which fail to fully capture the complexity of real materials. These models typically overlook intricate electronic structures and interactions that exist in actual systems, making experimental validation difficult.

In this paper, we introduce a numerical method to characterize the superconducting spectrum and topological invariants of two-dimensional (2D) slab systems based on first-principles calculations. This method is implemented within the open-source software WannierTools \cite{Wu2018wanniertools}, which allows for a more accurate description of the electronic structure of materials through Wannier functions. By bridging the gap between theoretical models and real material properties, this approach is expected to significantly enhance the alignment between theoretical predictions and experimental results in the study of TSCs.

The flow of the program is depicted in Fig.~\ref{program flow chart}. The process begins with the construction of Wannier functions from first-principles calculations, providing a precise representation of the electronic structure of the candidate material. Subsequently, s-wave superconducting pairing and Zeeman splitting are introduced at the interface of the 2D slab, simulating superconducting and magnetic proximity effects, respectively. This enables the construction of the corresponding 2D Bogoliubov-de Gennes (BdG) Hamiltonian, which is used to investigate the system’s superconducting properties.

Using this method, we explore a physical platform for realizing 2D chiral topological superconducting phases. Specifically, we focus on a heterostructure composed of a superconductor, candidate material, and ferromagnetic insulator. Our approach provides a robust framework for predicting and characterizing TSCs in experimentally accessible systems, facilitating further experimental investigation and potential technological applications.

\section{Methods}

\subsection{BdG Hamiltonian in slab system}

A widely used and efficient method for constructing tight-binding (TB) Hamiltonians for real materials is the maximally localized Wannier functions (MLWFs) \cite{marzari2012maximally}, as implemented in the Wannier90 software package \cite{mostofi2008wannier90}. This package provides interfaces to various first-principles calculation codes, including VASP, Wien2k, and Quantum ESPRESSO (QE), among others. As a result, MLWF Hamiltonians can be automatically generated from first-principles calculations~\cite{marzari2012maximally}, which give the real-space hopping parameters of electrons in real materials as:

\begin{equation}
  H_{mn}(\bm{R}) = \langle\mathcal{W}_{m\bm{0}}|\hat{H}_{KS}|\mathcal{W}_{n\bm{R}}\rangle
\end{equation}
where $\hat{H}_{KS}$ represents the Kohn-Sham (KS) Hamiltonian derived from first-principles calculations, $H_{mn}(R)$ denotes the projection of the KS Hamiltonian onto the MLWF basis, and $R$ indicates the lattice vector in the crystal structure. In the lattice gauge, the Fourier transform gives the bulk Hamiltonian:

\begin{equation}
  H_{mn}(\bm{k}) = \sum_{\bm{R}} H_{mn}(\bm{R}) e^{-i\bm{kR}}
\end{equation}

To investigate the superconducting proximity effects and magnetic proximity effects in superconductor/candidate material/ferromagnetic insulator heterojunctions, it is essential to employ open boundary conditions that effectively simulate the surface or interface of the material. For convenience, the three-dimensional vector $\bm{R}$ is divided into the longitudinal vector $\bm{R}_\perp$ and the transverse vector $\bm{R}_\parallel$, where $\bm{R}_\perp$ is perpendicular to the material surface and $\bm{R}_\parallel$ is parallel to it. By applying open boundary conditions in the $\bm{R}_\perp$ direction, a two-dimensional slab system is created, in which periodicity is maintained solely in the in-plane $\bm{R}_\parallel$ direction. However, it is important to note that the experimentally relevant dissociation plane may not align with the basis vectors defined by first-principles calculations. Consequently, the basis vectors $\bm{R}^{'}_{1,2,3}$ must be redefined to accommodate this discrepancy:

\begin{equation}
\begin{split}
   \bm{R}^{'}_1 = U_{11}\bm{R}_1+U_{12}\bm{R}_2+U_{13}\bm{R}_3 \\
   \bm{R}^{'}_2 = U_{21}\bm{R}_1+U_{22}\bm{R}_2+U_{23}\bm{R}_3 \\
   \bm{R}^{'}_3 = U_{31}\bm{R}_1+U_{32}\bm{R}_2+U_{33}\bm{R}_3 \\
\end{split}
\end{equation}
where $\bm{R}^{'}_1$ and $\bm{R}^{'}_2$ lattice vectors form $\bm{R}_\parallel$ and $\bm{R}^{'}_3$ corresponds to the $\bm{R}_\perp$ direction. With the basis-vector transformation matrix $U$, it is possible to generalize the physical properties of any dissociation surface.

The Hamiltonian for the two-dimensional (2D) thin film system can be derived by applying open boundary conditions in the $\bm{R}^{'}_3$ direction and subsequently performing a Fourier transform:

\begin{equation}
  H_{mn}^{slab}(\bm{k}_{\parallel}) = \sum_{\bm{R}^{'}_1,\bm{R}^{'}_2} H_{mn}(\bm{R}_\parallel^{'}) e^{-i\bm{k}_{\parallel}\bm{R}_\parallel^{'}}
\end{equation}

In this expression, the notation $\bm{R}_\parallel^{'}$ represents either $\bm{R}^{'}_1$ or $\bm{R}^{'}_2$. Different values of $\bm{R}^{'}_3$ correspond to distinct film layers, which constitute the longitudinal dimension of the 2D slab Hamiltonian. We denote the layer indices along the $\bm{R}^{'}_3$ as i and j.

For a system comprising $N_s$ layers, the matrix representation of the Hamiltonian for the 2D slab can be expressed as:

\begin{equation}
  H_{mn}^{slab}(\bm{k}_{\parallel})=\left[
  \begin{array}{cccc}
		H_{mn}^{11}(\bm{k}_{\parallel}) & H_{mn}^{12}(\bm{k}_{\parallel}) & \cdots & H_{mn}^{1N_s}(\bm{k}_{\parallel}) \\
		H_{mn}^{21}(\bm{k}_{\parallel}) & H_{mn}^{22}(\bm{k}_{\parallel}) & \cdots & H_{mn}^{2N_s}(\bm{k}_{\parallel}) \\
		\vdots & \vdots & \ddots & \vdots \\
		H_{mn}^{N_s1}(\bm{k}_{\parallel}) & H_{mn}^{N_s2}(\bm{k}_{\parallel}) & \cdots & H_{mn}^{N_sN_s}(\bm{k}_{\parallel})
  \end{array}
  \right]
\end{equation}

In this context, the diagonal matrix elements correspond to the intra-layer Hamiltonians, while the off-diagonal matrix elements represent the inter-layer Hamiltonians. The specific formulation of these matrix elements can be expressed as follows:

\begin{equation}
  H_{mn}^{ij}(\bm{k}_{\parallel}) = \sum_{\bm{R}^{'}=\{ \bm{R}^{'}_1,\bm{R}^{'}_2,(i-j)\bm{R}^{'}_3 \} } H_{mn}(\bm{R}^{'}) e^{-i\bm{k}_{\parallel}\bm{R}^{'}}
\end{equation}

In the presence of atomic-scale exchange coupling resulting from magnetic proximity effects, the Hamiltonian for the surface layer, defined by i = j = $N_s$, is expressed as follows:

\begin{equation}
  H_{mn}^{N_sN_s}(\bm{k}_{\parallel},\bm{M}) = H_{mn}^{N_sN_s}(\bm{k}_{\parallel})+\bm{M}\cdot\bm{\sigma}I_{orbital}
\end{equation}
where $\bm{M}$ represents the Zeeman field arising from the exchange coupling, $\bm{\sigma}$ denotes the Pauli matrices in spin space, and $I$ is the identity matrix in the Wannier orbital space.

To obtain the BdG Hamiltonian for the superconductor/candidate material/ferromagnetic insulator heterojunction, we induce superconducting pairing across the entire system, incorporating the effects of exchange coupling. In the Nambu representation, the BdG Hamiltonian can be expressed as follows:

\begin{equation}
  \label{BdG Hamiltonian}
  H_{BdG}^{slab}=\left[
  \begin{array}{cc}
		H_{mn}^{slab}(\bm{k}_{\parallel},\bm{M})-\mu & \Delta \\
		\Delta^{\dagger} & -H_{mn}^{slab}(-\bm{k}_{\parallel},\bm{M})^*+\mu
  \end{array}
  \right]
\end{equation}
where $\mu$ denotes the chemical potential of the system. Since the superconducting substrates used to realize topological superconductors are mostly s-wave superconductors, intra-orbital s-wave pairing, denoted by $\Delta$, is used throughout this paper.

\subsection{Superconducting proximity effect}

The superconducting gap induced by the proximity effect decays gradually with distance from the superconductor interface, a behavior that can be expressed as~\cite{silvert1975theory}:

\begin{equation}
\label{proximity effect}
\begin{aligned}	
\Delta_{\mathrm{decay}}&=\Delta_{\mathrm{SC}}\frac{\cosh[\alpha(z-d)]}{\cosh[\alpha d]} ~ , ~0<z<d \\
    \alpha^2&=\frac{t}{\Xi_0}(\frac{t}{\Xi_0}+\frac{3}{l}),
\end{aligned}
\end{equation}

\begin{figure}[t]
    \centering
    \includegraphics[width=0.5\textwidth]{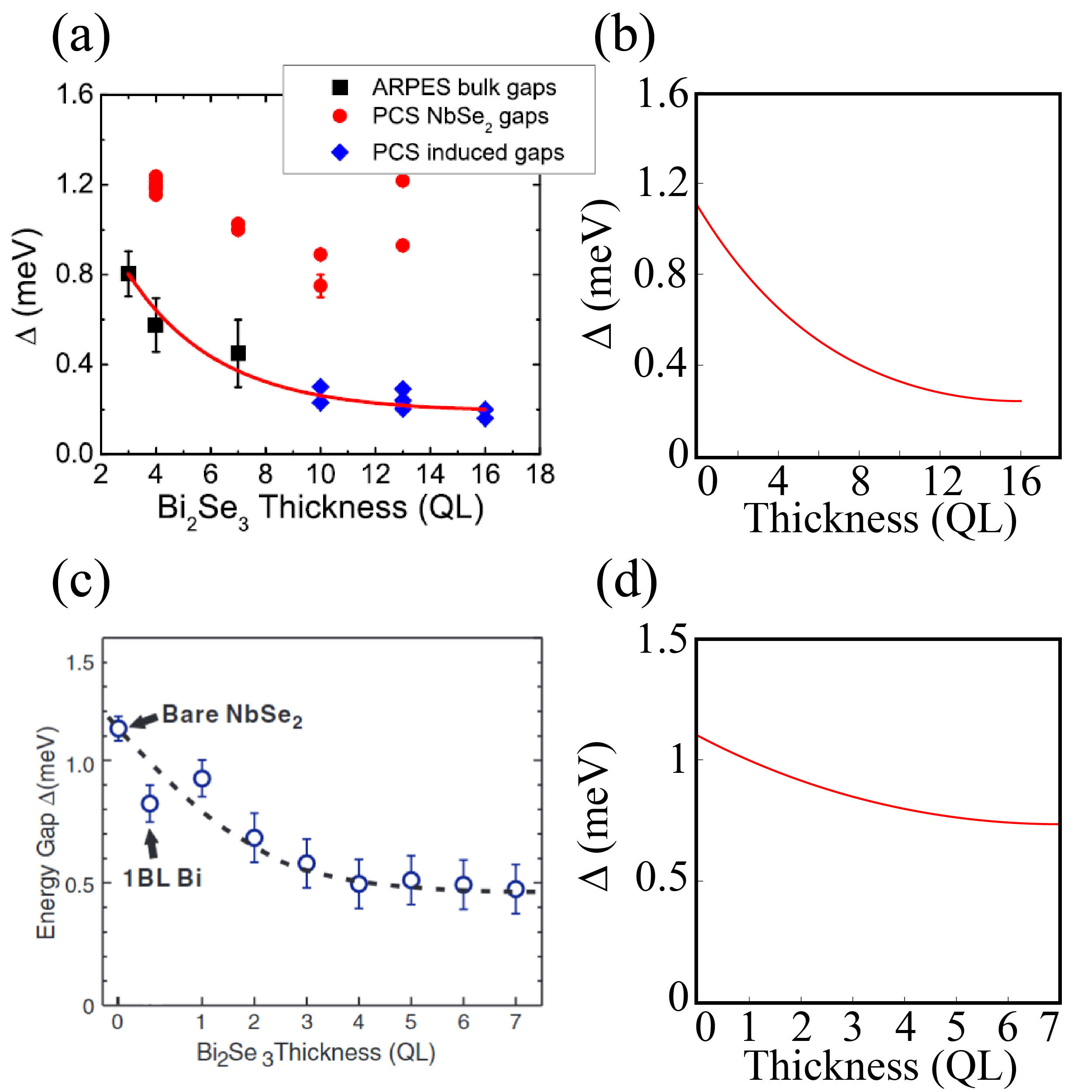}
    \caption{(a) and (c) are the experimental superconducting energy gap of different layers in reference~\cite{dai2017proximity} and~\cite{wang2012coexistence}. (b) and (d) are our fitting results. }
    \label{Bi2Se3 proximity effect}
\end{figure}

Let $z$ denote the distance from the surface of the SC, while $d$ represents the thickness of the candidate material. The parameter $t$ is defined as the ratio T/T$_c$, where T is the temperature and T$_c$ is the critical temperature. Additionally, $\Xi_0$ denotes the superconducting correlation length, and $l$ indicates the mean free path in the candidate material. We utilize these parameters to analyze the superconducting proximity effect in the NbSe$_2$/Bi$_2$Se$_3$ heterostructure. In Fig.\ref{Bi2Se3 proximity effect}, we demonstrate that this layer-dependent superconducting gap is consistent with the experimental findings of NbSe$_2$/Bi$_2$Se$_3$\cite{wang2012coexistence,dai2017proximity}, using the values $\Xi_0$=7.7nm~\cite{banerjee1997magnetic} and l=16nm~\cite{dai2017proximity}. Consequently, when accounting for the decay of superconducting pairing in heterostructures, $\Delta$ in Eq.~\ref{BdG Hamiltonian} should be replaced with $\Delta_{\mathrm{decay}}$.

\subsection{Superconducting topological invariant}

As summarized in Table~\ref{table class}, superconductors can be categorized into class D and class D$\rm\uppercase\expandafter{\romannumeral3}$ based on the presence or absence of time-reversal symmetry~\cite{altland1997nonstandard}. For two-dimensional (2D) systems, when time-reversal symmetry is broken, the system falls under the $\mathbb{Z}$ classification, with the corresponding topological invariant being the Chern number~\cite{thouless1982quantized}. Conversely, when time-reversal symmetry is preserved, the system is classified as $\mathbb{Z}_2$, where the topological invariant is the $\mathbb{Z}_2$topological number defined by Kane and Mele~\cite{kane2005z}. Furthermore, Qi et al. demonstrated that in a superconducting system possessing both time-reversal and inversion symmetries, s-wave superconducting pairing results in a topologically trivial phase~\cite{qi2010topological}.

\begin{table}[ht]
  \begin{center}
    \caption{Time-reversal-breaking (TRB) and time-reversal-invariant (TRI) Topological periodic table of superconductors. In the Altland-Zirnbauer classification~\cite{altland1997nonstandard}, they belong to categories D and D$\rm\uppercase\expandafter{\romannumeral3}$, respectively.}
    \begin{tabular}{l|c|c|c|c|r}
      \label{table class}
      AZ class & TRS & PHS & 1D & 2D & 3D \\
      \hline
      TRB SCs (class D) & --- & +1 & $\mathbb{Z}_2$ & $\mathbb{Z}$ & 0 \\
      TRI SCs (class D$\rm\uppercase\expandafter{\romannumeral3}$) & -1 & +1 & $\mathbb{Z}_2$ & $\mathbb{Z}_2$ & $\mathbb{Z}$ \\
    \end{tabular}
  \end{center}
\end{table}

There are many methods for calculating topological invariants~\cite{yu2011equivalent,soluyanov2011computing,fukui2007quantum,fu2007topological}, among which the Wannier charge centers (Wilson loop) method~\cite{yu2011equivalent,soluyanov2011computing} can calculate the $Z_2$ topological number of a system as well as the Chern number. In the following, the topological invariants of the BdG Hamiltonian are calculated using the Wannier charge centers (Wilson loop) method in the 2D slab system. In WannierTools, we take the algorithm presented in Refs~\cite{yu2011equivalent,gresch2017z2pack}. The hybrid Wannier functions~\cite{coh2009electric} are defined as:

\begin{equation}
  |nk_xl_y\rangle = \frac{1}{2\pi}\int^{2\pi}_0 dk_y e^{-ik_yl_y}|\psi_{n\bm{k}_{\parallel}}^{BdG}\rangle
\end{equation}
where $|\psi_{n\bm{k}_{\parallel}}^{BdG}\rangle$ is the Bloch wave function of BdG Hamiltonian Eq.~\ref{BdG Hamiltonian}, $k_x$ and $k_y$ represent the reciprocal lattice vectors in the x and y directions of the 2D plane formed by $\bm{k}_{\parallel}$. The hybrid Wannier centers are defined as:

\begin{equation}
\label{hybrid Wannier centers}
\begin{split}
  \bar{y}_n(k_x) &= \langle nk_x0|y|nk_x0\rangle \\
  &= \frac{i}{2\pi}\int^{\pi}_{-\pi}dk_y\langle u_{n,k_x,k_y}^{BdG}|\partial_{k_y}|u_{n,k_x,k_y}^{BdG}\rangle
\end{split}
\end{equation}
where $|u_{n,k_x,k_y}^{BdG}\rangle$ is the periodic part of Bloch function $|\psi_{n\bm{k}_{\parallel}}^{BdG}\rangle$. In practice, the integration over $k_y$ is transformed by a summation over the
discretized $k_y$. Eq.~\ref{hybrid Wannier centers} can be reformulated using the discretized Berry phase formula~\cite{marzari1997maximally}:

\begin{equation}
\begin{split}
  \bar{y}_n(k_x) = \frac{1}{2\pi} \mathrm{Im}\ln\prod_j M_{nn}^{(j)}
\end{split}
\end{equation}
where the gauge-dependent overlap matrix $M_{mn}^{(j)}=\langle u_{m,k_x,k_{y_i}}^{BdG}|u_{n,k_x,k_{y_{i+1}}}^{BdG}\rangle$ is introduced. The method for calculating $\bar{y}_n(k_x)$ is well-established~\cite{marzari1997maximally,king1993theory,soluyanov2011wannier,yu2011equivalent,Wu2018wanniertools}, we can get the topological properties of a 2D system from the evolution of $\bar{y}_n(k_x)$ along a $k_x$ string. The details of such classification of Wannier charge centers or Wilson loop are discussed in Refs.~\cite{yu2011equivalent,soluyanov2011computing,soluyanov2011wannier}. More information can be found in Ref.~\cite{gresch2017z2pack}.

In our program, the atomic positions in the slab system are given in Cartesian coordinates, and the positions of the particles and holes are assumed to correspond in a one-to-one manner. Additionally, the number of occupied states in the BdG spectrum is always half of the total number of BdG spectrum, meaning that the Wilson loop is computed for half of the total BdG spectrum.

\subsection{Proposals to induce TSC by s-wave SC}
\label{Proposals}

\subsubsection{Fu-Kane's proposal}

Due to the superconducting proximity effect, the Bogoliubov-de Gennes (BdG) Hamiltonian at the interface of a 3D strong topological insulator (TI) and an s-wave superconductor can be written as~\cite{fu2008superconducting}:

\begin{equation}
  H_{interface}(k)=k_x\tau_0\sigma_x+k_y\tau_z\sigma_y-\mu\tau_z\sigma_0+\Delta\tau_y\sigma_y\label{hinterface}
\end{equation}

This Hamiltonian is written in the Nambu basis $(c_{\uparrow,k}^{\dagger},c_{\downarrow,k}^{\dagger},c_{\uparrow,-k},c_{\downarrow,-k})$, where $\tau$ and $\sigma$ represent the Pauli matrices of particle-hole space and spin space respectively, $\mu$ represents the chemical potential and $\Delta$ represents the superconducting pairing strength. For $\mu\gg\Delta$, take the gauge transformation $a_k=c_{\uparrow,k}+e^{i\theta_k}c_{\downarrow,k}$ for this Hamiltonian (Eq.\ref{hinterface}), where $\theta_k$ is defined by $|k|(cos\theta_k,sin\theta_k)=(k_x,k_y)$, an effective $p+ip$-wave pairing can be obtained:

\begin{equation}
  H_{eff}^{p_x+ip_y}(k)=\left[
  \begin{array}{cccc}
		-|k|-\mu & 0 & 0 & -\Delta e^{i\theta_k} \\
		0  & |k|-\mu & \Delta e^{i\theta_k} & 0 \\
		0  & \Delta e^{-i\theta_k} & -|k|+\mu & 0 \\
		-\Delta e^{-i\theta_k} & 0 & 0 & |k|+\mu
  \end{array}
  \right]
\end{equation}

The physical interpretation of this process involves treating $\Delta$ as a perturbation and projecting the Hamiltonian onto the eigenstates of the unperturbed system. However, a key distinction between this Hamiltonian and that of $p+ip$-wave superconductors is that the former retains time-reversal symmetry. As a result, the chiral boundary states characteristic of $p+ip$-wave superconductivity are prohibited in this system. When the Zeeman field $M_z\tau_z\sigma_z$ is introduced to the system, it breaks time-reversal symmetry. In this context, it is evident that topological phases can emerge in systems belonging to class D, and these topological phases can be characterized by the Chern number~\cite{sato2017topological}. The condition for the phase transition is that $|M_z|>\sqrt{\mu^2+\Delta^2}$. Through the analysis above, it is evident that two-dimensional systems with Rashba spin-orbit coupling (SOC) exhibit a situation very similar to that described here. By introducing a Zeeman field to break time-reversal symmetry, topological superconductivity can be achieved~\cite{kezilebieke2020topological}.

\subsubsection{Nagaosa's proposal}

Consider a superconductor-TI film-ferromagnet heterostructure as illustrated in Fig.\ref{fig1}~\cite{he2019platform}. The low-energy physics of this system is governed by the two Dirac surface states of the topological insulator. Accordingly, the BdG Hamiltonian of the system in the Nambu basis $(c_{t\uparrow,k}^{\dagger},c_{t\downarrow,k}^{\dagger},c_{b\uparrow,k}^{\dagger},c_{b\downarrow,k}^{\dagger},c_{t\uparrow,-k},c_{t\downarrow,-k},c_{b\uparrow,-k},c_{b\downarrow,-k})$ is given by:

\begin{figure}[t]
    \centering
    \includegraphics[width=0.5\textwidth]{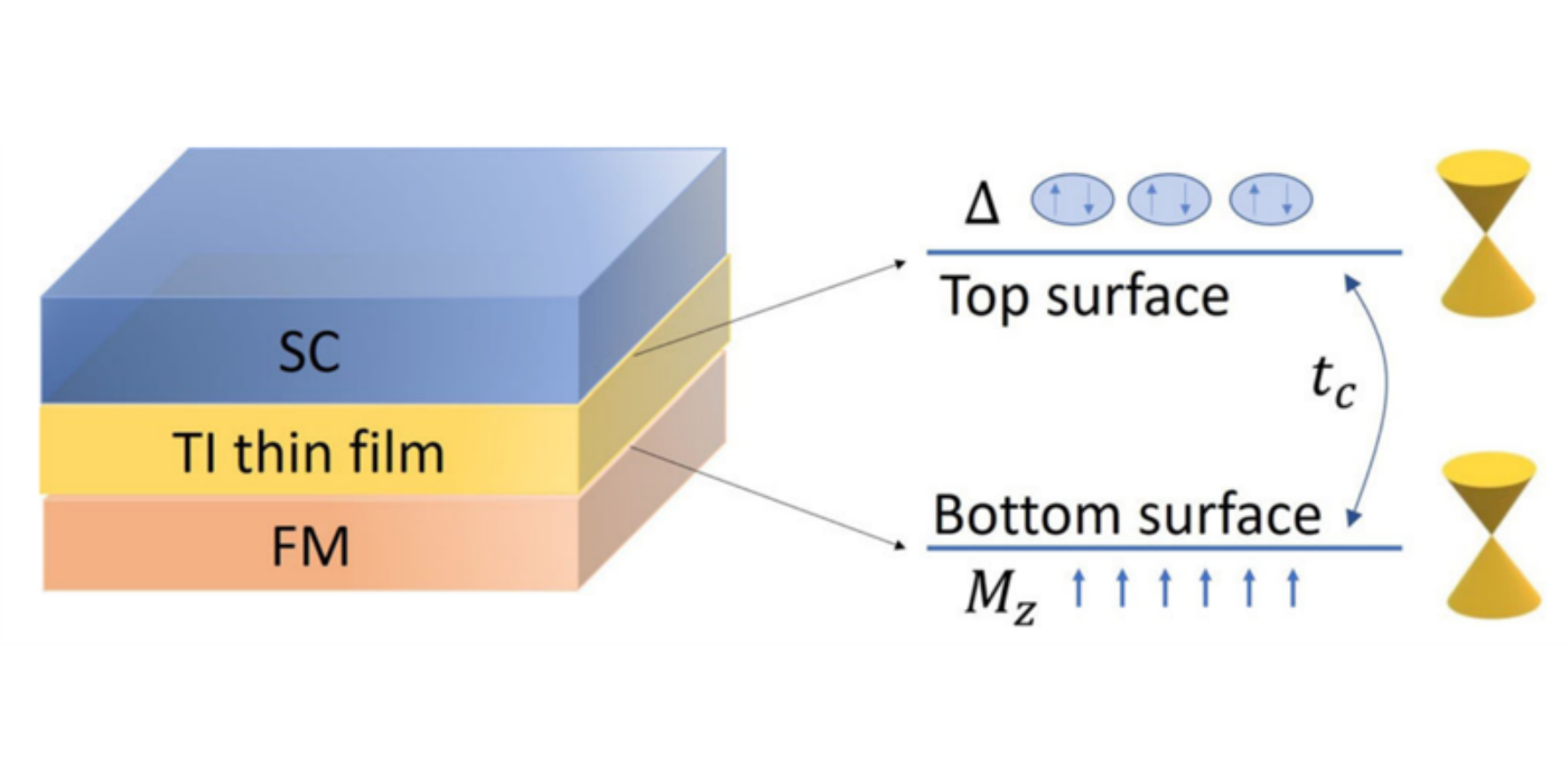}
    \caption{A sketch plot of Nagaosa's proposal to realize a topological superconductor.}
    \label{fig1}
\end{figure}

\begin{equation}
\begin{split}
  H_{twosurfaces}(k)=&v(k_x\sigma_y\tau_z-k_y\sigma_x\tau_0)s_z+t_c\sigma_0 s_x\tau_z\\
  &+M_z\sigma_z\tau_z\frac{s_z-s_0}{2}+\Delta \sigma_y\tau_y\frac{s_z+s_0}{2}\\
  &-\sigma_0(\mu s_0+\delta E s_z)\tau_z
\end{split}
\end{equation}
$\tau$, $\sigma$, and $s$ represent the Pauli matrices of particle-hole space, spin space, top and bottom space, respectively. $M_z$ represents the Zeeman field on the bottom surface, $\Delta$ represents the s-wave superconducting pairing strength on the top surface, $t_c$ represents the coupling between the top and bottom surfaces, $\mu$ represents the chemical potential, and $\delta E$ represents the energy difference between the top and bottom surfaces. This Hamiltonian describes the interface between two Dirac cones characterized by distinct mass terms: $\Delta$ on the top and $M_z$ on the bottom. According to the criteria established in~\cite{bernevig2013topological}, this configuration corresponds to a chiral topological superconducting state.

Let's briefly analyze the topological properties of this system. In the limit where both $t_c$ and $\delta E$ approach zero, the Hamiltonian $H_{twosurfaces}(k)$  for the top surface can be expressed as shown in Eq.~\ref{hinterface}. As discussed in the previous section, we conclude that the top surface is topologically trivial. The Hamiltonian for the bottom surface can be written as:

\begin{equation}
  H_{bottom}(k)=k_x\tau_0\sigma_x+k_y\tau_z\sigma_y-\mu\tau_z\sigma_0+M_z\tau_z\sigma_z
\end{equation}

This Hamiltonian is also block diagonal, corresponding to two Dirac Hamiltonians with opposite helicity and opposite masses. Specifically, if $M_z$ is not zero, the system must exhibit topological characteristics,  which are determined by the mass term ($C=\frac{1}{2}sign(M_z)$ for each Dirac Hamiltonian). Moreover, if the superconducting proximity effect can extend to the bottom surface, then, analogous to the scenario discussed in the previous section, the system enters a topological superconducting state when the condition $|M_z|>\sqrt{\mu^2+\Delta^2}$ is satisfied.

Our program is based on the heterostructure depicted in Fig.~\ref{fig1}, where both the superconducting proximity effect and the magnetic proximity effect are introduced in the candidate materials situated in the middle layer. Ultimately, the topological superconducting properties of the system are calculated using the Hamiltonian in the form presented in Eq.~\ref{BdG Hamiltonian}. In the following sections, we will validate the reliability of our approach through model calculations as well as computations on real materials, such as Bi$_2$Se$_3$ and SnTe.

\section{Examples}

\subsection{TI model}

\begin{figure*}[ht]
    \centering
    \includegraphics[width=0.95\textwidth]{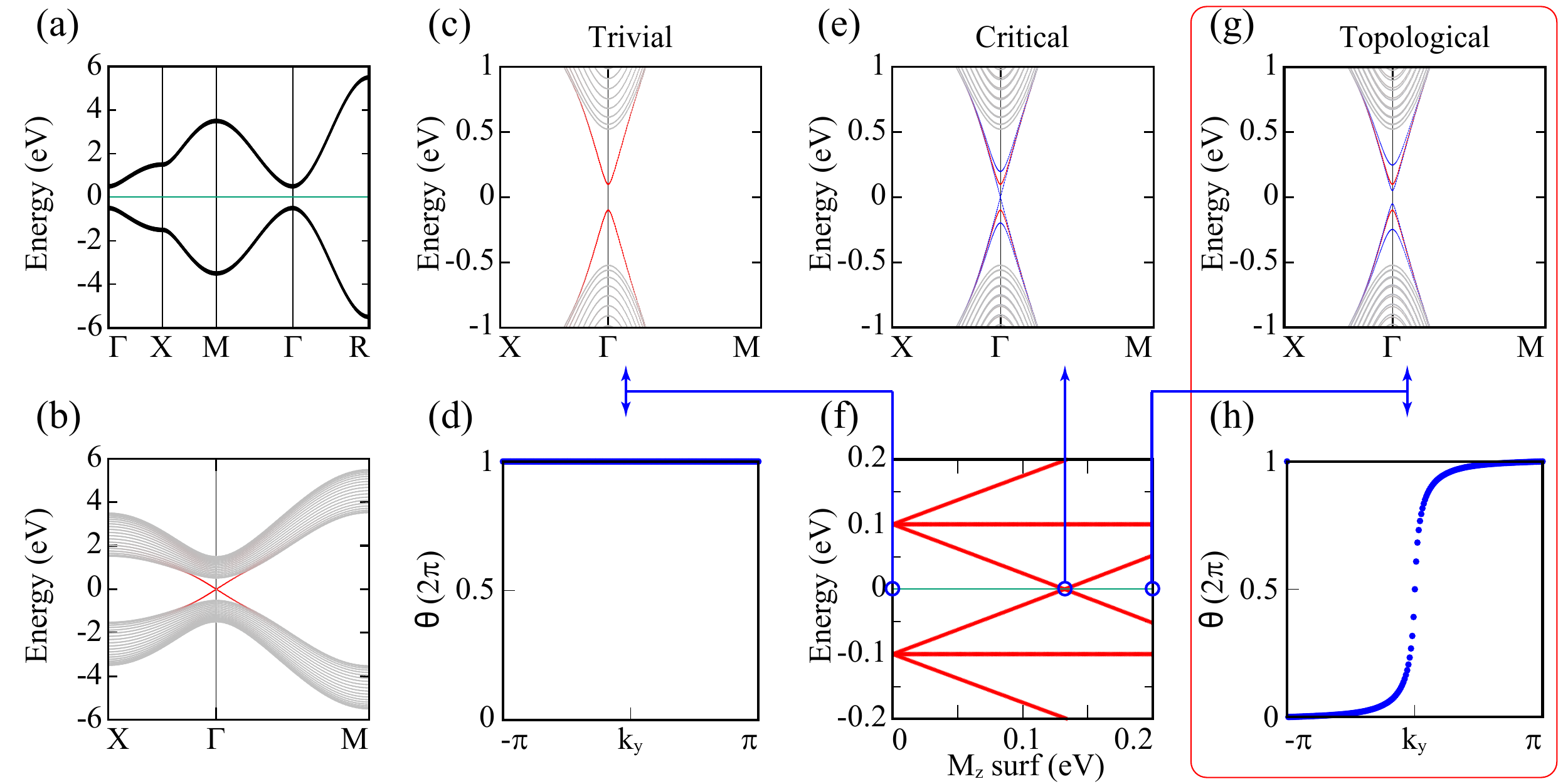}
    \caption{a: Band structure corresponding to the TI model. b: Band structure of the 20-layer slab in the (001) direction. c, d: BdG spectrum, and the Wilson loop with $M_z=0$, the corresponding Chern number is 0. e: BdG spectrum for $M_z=0.132$, the system is at the critical point of a topological phase transition. f: Energy of the BdG Hamiltonian of the system at $\Gamma$ point as a function of the surface Zeeman splitting $M_z$, This can be viewed as a topological phase diagram. g, h: BdG spectrum, and the Wilson loop with $M_z=0.2$, the corresponding Chern number is 1. (In panels c, e, and g, the red and blue colors represent the top and bottom surface states, respectively.)}
    \label{TI}
\end{figure*}

Before conducting calculations on real materials, we first perform numerical simulations and analyze the BdG Hamiltonian of tight-binding (TB) model for the topological insulator. These numerical results are instrumental in establishing a clear physical understanding of the system and serve to validate the accuracy of our computational implementation. The model is written on a simple cubic lattice with two orbitals per primitive cell:

\begin{equation}
  H_{\mathrm{TI}}(k)=m(k)\tau_z\sigma_0+\tau_x\bm{d(k)}\cdot\bm{\sigma}
\end{equation}
where $\tau$ ($\sigma$) are Pauli matrices in the orbital (spin) space, $m(k)=M+m_0\sum_i\cos(k_i),d(k)^i=2t\sin{k_i},i=x,y,z$, and $t$, $m_0$, and $M$ are parameters of the model. The model is in the strong TI phase if $-3<\frac{M}{m_0}<-1$. In our calculations, we employed the parameters $t=0.5$, $M=2.5$, $m_0=-1$. The resulting bulk band spectrum is presented in Fig.~\ref{TI} a. To investigate the surface states, we constructed a two-dimensional slab system consisting of 20 layers, with the boundary opened in the (001) direction. The calculated band spectrum for this slab system is shown in Fig.~\ref{TI} b.

This section examines the topological superconducting properties of the slab Hamiltonian by substituting it into Eq.~\ref{BdG Hamiltonian}, while taking into account the magnetic proximity effect and superconducting pairing. In our model calculations, we temporarily neglect the decay of the superconducting gap by assuming a temperature of zero. We introduce a uniform superconducting pairing $\Delta=0.1$ to the slab system. With the chemical potential set to $\mu=0$ and surface Zeeman splitting $M_z=0$, the BdG spectrum of the system exhibits a superconducting gap and remains in a topologically trivial phase, as illustrated in Fig.~\ref{TI} c, d.

Subsequently, we utilize the phase diagram plotting function to calculate the BdG spectrum at the $\Gamma$ as a function of the Zeeman splitting. The results are presented in Fig.~\ref{TI} f. It is observed that the BdG spectrum reaches a critical point of phase transition when $M_z$ equals 0.132, as shown in Fig.~\ref{TI} e. Upon exceeding this value, the system transitions into a topological superconducting state characterized by a Chern number of 1. The corresponding BdG spectrum and Wilson loop for this state are shown in Fig.~\ref{TI} g, h.

From the preceding analysis, the physical mechanism underlying the emergence of topological superconducting states is elucidated. The system transitions into a topological superconducting phase whenever the Zeeman field $M_z$ induces a closing and reopening of the BdG spectrum. This principle remains consistent across the examples discussed.

\subsection{Bi$_2$Se$_3$}

\begin{figure*}[ht]
    \centering
    \includegraphics[width=0.95\textwidth]{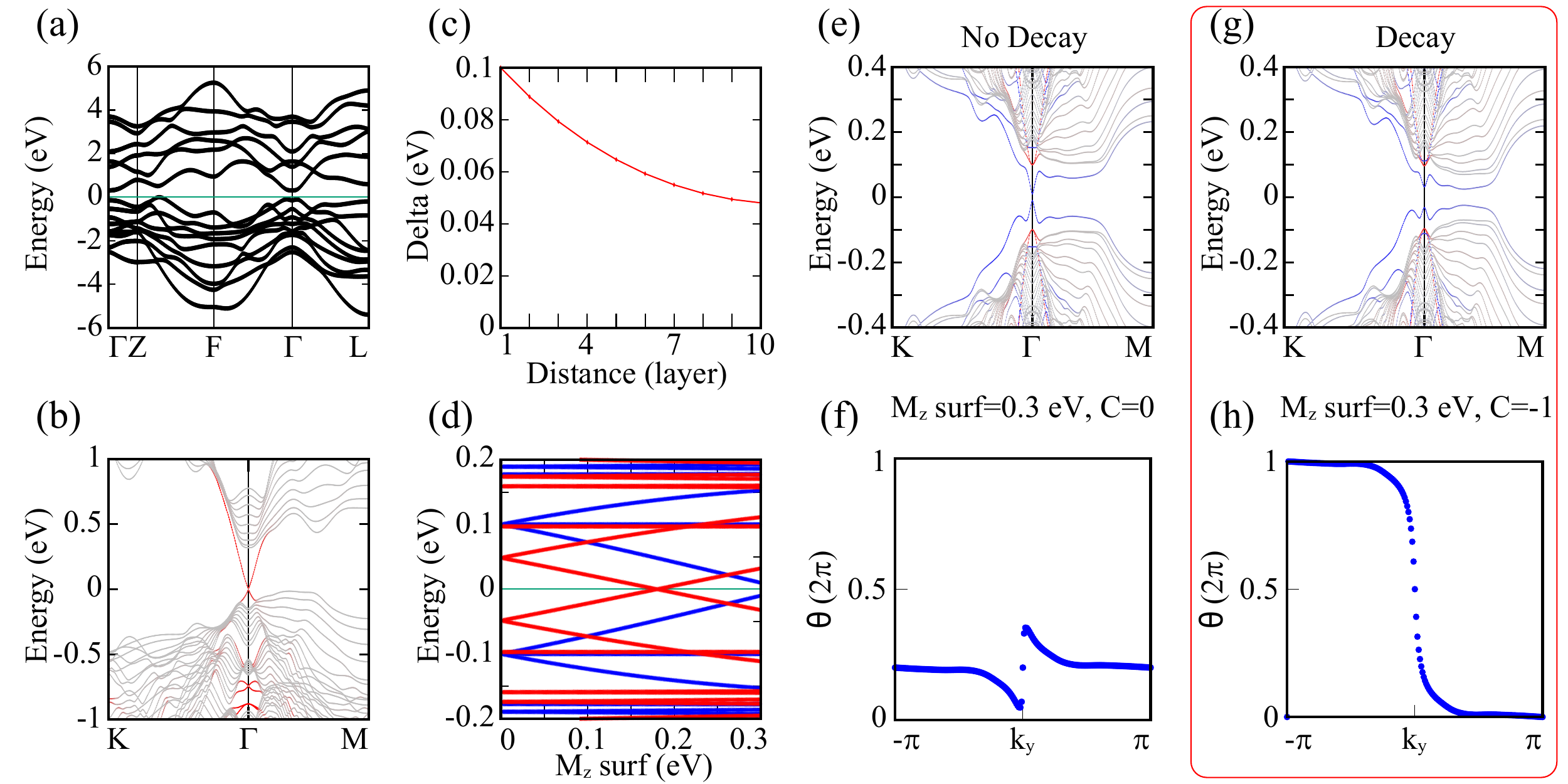}
    \caption{a: Band structure corresponding to Bi$_2$Se$_3$. b: Band structure of the 10-layer slab in the (001) direction. c: The decay of superconducting pairing with the number of layers. d: Energy of the BdG Hamiltonian of the system at $\Gamma$ point as a function of the surface Zeeman splitting $M_z$, This can be viewed as a topological phase diagram. The blue lines represent the phase diagrams for superconducting pairing that is uniformly distributed without decay, while the red lines correspond to the phase diagrams when decay is taken into account. e, f: BdG spectrum, and the Wilson loop with $M_z=0.3 eV$ when decay is not considered. It is a topologically trivial phase. g, h: BdG spectrum, and the Wilson loop with $M_z=0.3 eV$ when decay is considered, the corresponding Chern number is -1.}
    \label{Bi2Se3}
\end{figure*}

In this section, we demonstrate the program's capability to analyze topological superconductivity in real material systems, using Bi$_2$Se$_3$ as a representative example. For the calculations involving real materials, we aim to emulate the actual heterogeneous structure depicted in Fig.~\ref{fig1}. To achieve this, we incorporate the spatially decaying superconducting pairing from Eq.~\ref{proximity effect} and the short-range exchange interaction at the ferromagnet interface. Our goal is to induce a two-dimensional chiral topological superconducting state at the interface between the candidate material and the ferromagnet. Bi$_2$Se$_3$, a topological insulator, serves as our testbed. Utilizing first-principles calculations, we derived the TB model for the p-orbitals of Bi and Se atoms using Wannier90. The bulk band spectrum and the 10-layer slab band spectrum in the (001) direction, obtained from this TB model, are presented in Fig.~\ref{Bi2Se3} a and b respectively, with the Dirac cone emerging in the gap~\cite{zhang2009topological}.

By substituting the TB Hamiltonian of Bi$_2$Se$_3$ into Eq.~\ref{BdG Hamiltonian} and considering Ns=10, we derive the BdG Hamiltonian for a 10-layer Bi$_2$Se$_3$ system, taking into account the effects of superconducting and magnetic proximity. In this analysis, we employ the decaying form described in Eq.~\ref{proximity effect},  with the following parameters: $\Delta_{\mathrm{SC}}=0.1eV$, t=0.6, $\Xi_0$=7.7nm~\cite{banerjee1997magnetic}, l=16nm~\cite{dai2017proximity}. The dependence of the superconducting pairing on the number of layers is illustrated in  Fig.~\ref{Bi2Se3} c.

Building upon the analysis presented above, we can construct the topological phase diagram of the system, as depicted in Fig.~\ref{Bi2Se3} d. In this diagram, the blue lines represent the phase boundaries obtained without considering decay effects, while the red lines indicate the phase boundaries when decay is taken into account. For a fixed parameter of $M_z=0.3 eV$, the BdG spectrum and Wilson loop corresponding to the case without decay are illustrated in Fig.~\ref{Bi2Se3} e, f, which reveal a topologically trivial phase. Conversely, when decay effects are included, the BdG spectrum and Wilson loop are presented in Fig.~\ref{Bi2Se3} g, h, indicating the emergence of a topological superconducting state characterized by a Chern number of -1. These results suggest that incorporating the decay of $\Delta$ facilitates the realization of the topological superconducting phase at smaller Zeeman fields, aligning more closely with the behavior observed in actual materials.

\subsection{SnTe}

\begin{figure*}[ht]
    \centering
    \includegraphics[width=0.95\textwidth]{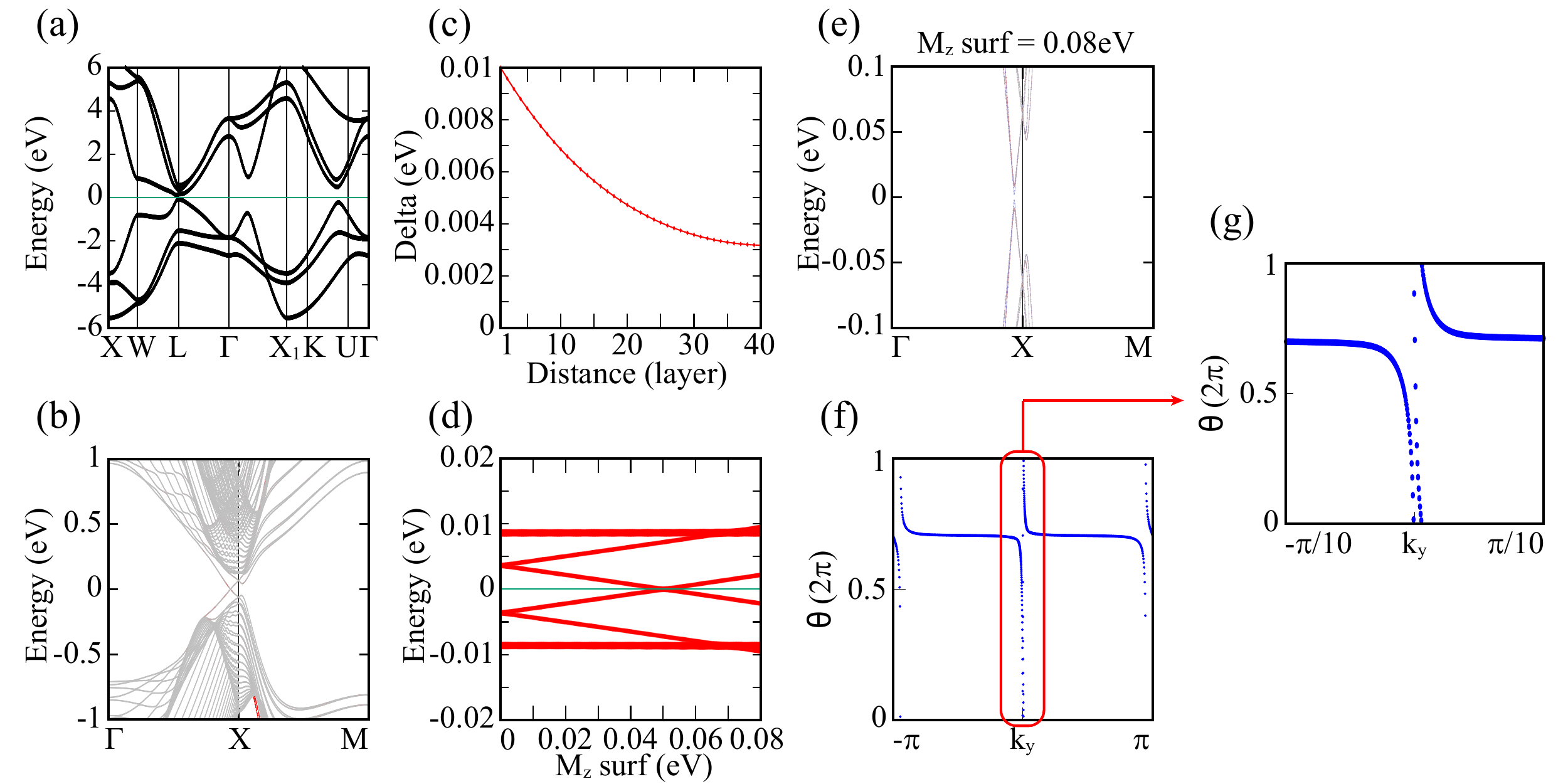}
    \caption{a: Band structure corresponding to SnTe. b: Band structure of the 40-layer slab in the (001) direction. c: The decay of superconducting pairing with the number of layers. d: Energy of the BdG Hamiltonian of the system at surface Dirac cone as a function of the surface Zeeman splitting $M_z$, This can be viewed as a topological phase diagram. e, f: BdG spectrum, and the Wilson loop with $M_z=0.08 eV$, the corresponding Chern number is -4. g: partial enlargement of f to see that the Chern number of the system at this point is -4.}
    \label{SnTe}
\end{figure*}

Based on the calculations presented in the previous two subsections, we have established that Dirac cone surface states provide a means to construct 2D chiral topological superconductors. In the following, we investigate whether systems with multiple surface Dirac cones can support 2D topological superconductors with high Chern numbers, using SnTe as a representative example.

SnTe is a topological crystal insulator protected by mirror symmetry, with four Dirac cone surface states on its (001) surface~\cite{hsieh2012topological}. Using first-principles calculations, we constructed a TB model for the p-orbitals of Sn and Te atoms, implemented through the Wannier90 software. The resulting bulk band structure and the band spectrum of a 40-layer slab along the (001) direction, derived from this TB model, are presented in Fig.~\ref{SnTe} a, b, respectively.

Here, we adopt the decaying form of Eq.~\ref{proximity effect}, with a superconducting gap $\Delta_{\mathrm{SC}}=0.01eV$, while keeping the other parameter values consistent with those used for Bi$_2$Se$_3$. The decay of the superconducting pairing as a function of the number of layers is shown in Fig.~\ref{SnTe} c.

Next, we derived the topological phase diagram of the system, as shown in Fig.~\ref{SnTe} d. The BdG spectrum and Wilson loop for $M_z=0.08 eV$ are shown in Fig.~\ref{SnTe} e, f. Due to the requirement of incorporating both Dirac cones within the same integration path for the Wilson loop calculations, the Wilson loop spectrum of SnTe exhibits a pronounced gradient in the region where the Dirac cones overlap. By focusing on the Wilson loop spectrum near the $\Gamma$ point, we can clearly identify that the Chern number of the system is -4, as shown in Fig.~\ref{SnTe} g.

This result is expected, as the system lacks time-reversal symmetry, and the four surface Dirac cones are related solely by a fourfold rotational symmetry. Consequently, these four Dirac cones should each contribute an equal Berry curvature, resulting in Chern number being a multiple of four.

\subsection{NbSe$_2$ model}

\begin{figure*}[ht]
    \centering
    \includegraphics[width=0.95\textwidth]{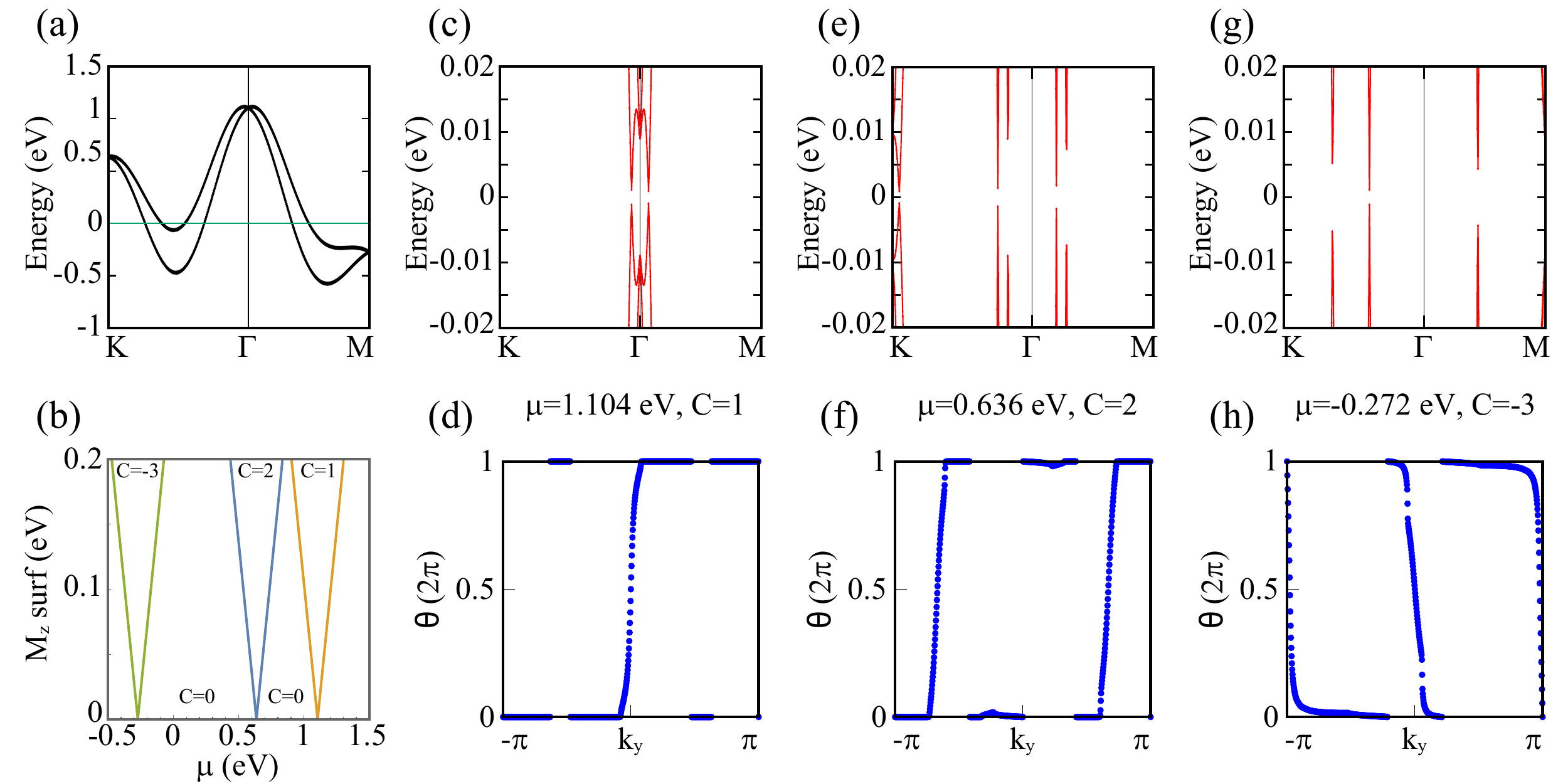}
    \caption{a: Band structure corresponding to the NbSe$_2$ model. b: The phase diagram of the whole system with respect to B$_z$ and $\mu$. c, d: BdG band spectrum, and the Wilson loop with $\mu=1.104 eV$, the corresponding Chern number is 1. e, f: BdG band spectrum, and the Wilson loop with $\mu=0.636 eV$, the corresponding Chern number is 2. g, h: BdG band spectrum, and the Wilson loop with $\mu=-0.272 eV$, the corresponding Chern number is -3.}
    \label{NbSe2}
\end{figure*}

In this section, we extend the realization scheme of topological superconductivity from topological insulators to quasi-two-dimensional materials with a Rashba SOC. We begin with the TB model of monolayer NbSe$_2$ and replicate the findings presented in~\cite{kezilebieke2020topological} using our approach. The electronic bands of NbSe$_2$ near the Fermi energy are primarily derived from Nb d-orbitals, and the relevant low-energy physics can be qualitatively captured by a minimal model that incorporates one effective d orbital per site on a triangular lattice.

\begin{equation}
  H_{\mathrm{NbSe_2}}(k)=E_t(k)\sigma_0+E_{R_x}(k)\sigma_x+E_{R_y}(k)\sigma_y
\end{equation}

We consider a simple third-neighbor hopping form for the normal band structure:

\begin{equation}
\begin{split}
  E_t(k)= &-2t_1[\cos{k_xa}+2\cos{\frac{k_xa}{2}}\cos{\frac{k_ya\sqrt{3}}{2}}]\\ &-2t_2[\cos{k_y\sqrt{3}a}+2\cos{\frac{k_x3a}{2}}\cos{\frac{k_y\sqrt{3}a}{2}}]\\
  &-2t_3[\cos{2k_xa}+2\cos{k_xa}\cos{k_y\sqrt{3}a}]
\end{split}
\end{equation}

The SOC terms are given by:

\begin{equation}
\begin{split}
  &E_{R_x}(k)= 2\alpha_R\sqrt{3}\sin{\frac{k_ya\sqrt{3}}{2}\cos{\frac{k_xa}{2}}}\\
  &E_{R_y}(k)= -2\alpha_R(\sin{k_xa}+\sin{\frac{k_xa}{2}}\cos{\frac{k_ya\sqrt{3}}{2}})
\end{split}
\end{equation}

For the subsequent calculations, we employed the parameters from~\cite{kezilebieke2020topological}: $t_1=-0.04 eV$, $t_2=-0.132 eV$, $t_3=-0.012 eV$, $\alpha_R=0.06 eV$. The band structure of the model is shown in Fig.\ref{NbSe2} a.

Next, we set $\Delta=0.001 eV$, adjust the chemical potential, and utilize our phase diagram function to plot the phase transition points at various chemical potentials. By connecting these points, we obtain the phase diagram of the system as a function of B$_z$ and $\mu$, as shown in Fig.~\ref{NbSe2} b. It is evident that this phase diagram is in perfect agreement with the one presented in~\cite{kezilebieke2020topological}.

Finally, we computed the Chern numbers for different regions of the phase diagram by setting B$_z=0.01 eV$. We identified three topologically non-trivial regions: the region with a Chern number of C=1, as shown in Fig.~\ref{NbSe2} c, d; the region with C=2, as shown in Fig.~\ref{NbSe2} e, f; and the region with C=-3, as shown in Fig.~\ref{NbSe2} g, h.

As discussed in~\cite{kezilebieke2020topological}, the absolute values of the different Chern numbers can be explained qualitatively. When the chemical potential intersects the band degeneracies at the three high-symmetry points shown in  Fig.\ref{NbSe2} a,  the Berry fluxes at these points contribute to the generation of distinct Chern numbers. In the first Brillouin zone, there is one $\Gamma$ point, three M points, and two equivalent K and K$^{'}$ points, so three different topological phases are produced.

\subsection{InSb}

\begin{figure*}[ht]
    \centering
    \includegraphics[width=0.95\textwidth]{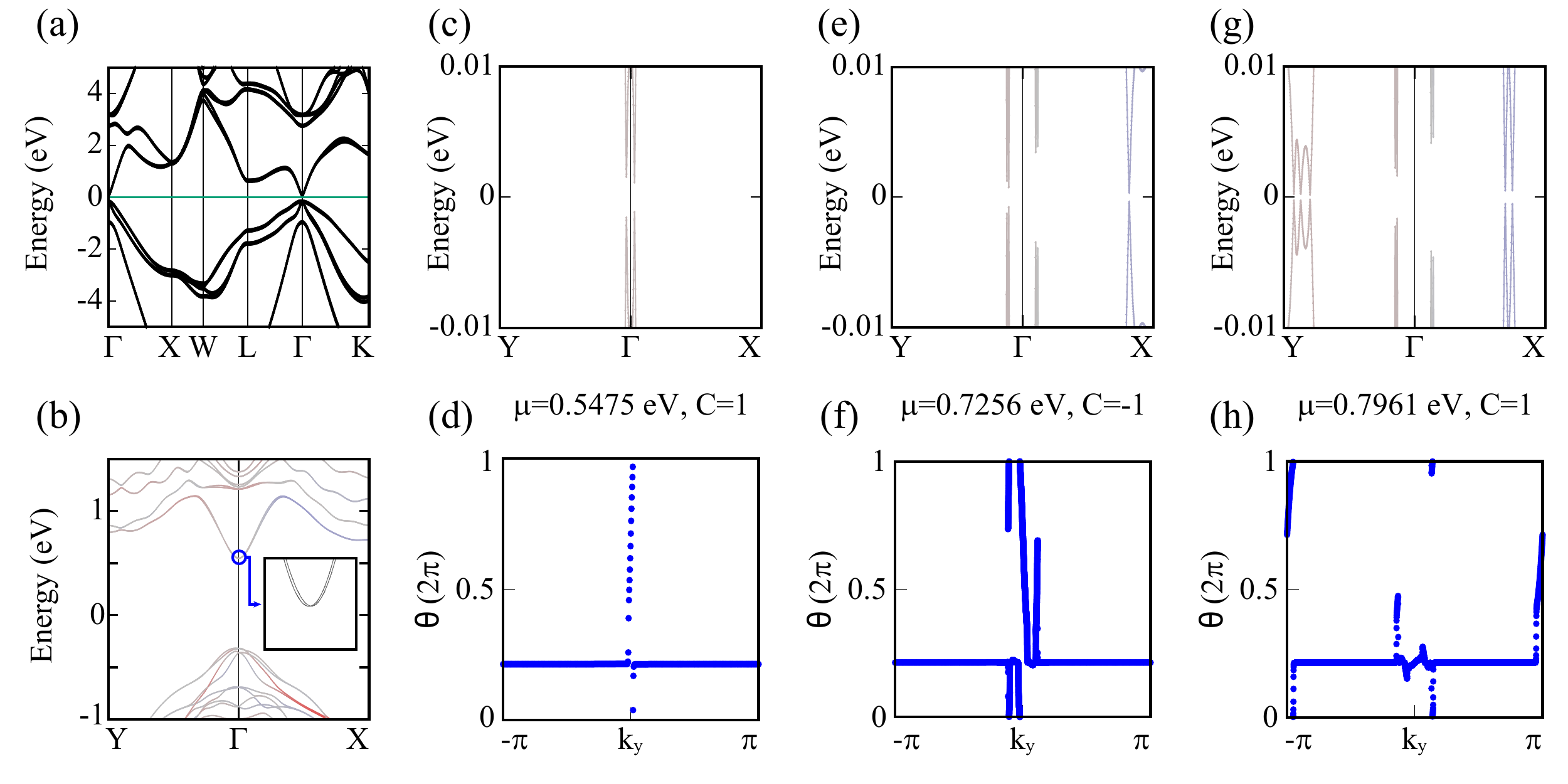}
    \caption{a: Band structure corresponding to the TB model of InSb. b: Band structure of the 10-layer slab in the (001) direction. c, d: BdG band spectrum, and the Wilson loop with $\mu=0.5475 eV$, the corresponding Chern number is 1. e, f: BdG band spectrum, and the Wilson loop with $\mu=0.7256 eV$, the corresponding Chern number is -1. g, h: BdG band spectrum, and the Wilson loop with $\mu=0.7961 eV$, the corresponding Chern number is 1.}
    \label{InSb}
\end{figure*}

Building on the previous example, we have established that topological superconducting states can be realized in 2D systems with Rashba SOC. In this section, we examine indium antimonide (InSb), a material commonly used for the fabrication of Majorana nanowires, to demonstrate the feasibility of our proposed scheme in real materials.

The crystal structure of InSb adopts the zincblende form, which is equivalent to a face-centered cubic (FCC) lattice with one formula unit per primitive unit cell. Following the approach outlined in~\cite{soluyanov2016optimizing}, we employed HSE06 hybrid functionals for first-principles calculations and constructed a TB model for InSb using the s and p orbitals of In and the p orbitals of Sb. The band structure derived from this TB model is shown in Fig.~\ref{InSb} a and is in good agreement with the results presented in~\cite{soluyanov2016optimizing}.

Next, we calculated the band structure of a 10-layer slab InSb along the (001) direction. To mitigate the effects of dangling bonds, we adopted the method described in~\cite{soluyanov2016optimizing}. Specifically, we modified the on-site energies of the s and $p_x$ orbitals of In atoms on the bottom surface by +5 eV, and the energy of $p_y$ and $p_z$ orbitals of Sb atoms on the top surface by -5 eV. The resulting modified band structure is shown in Fig.~\ref{InSb} b.

Due to the extensive range of parameter regimes in which topological superconducting phases can be realized in real materials, we do not present a complete topological phase diagram here. Instead, we focus on the topological superconducting states derived from the lowest conduction bands at the $\Gamma$, X, and Y points, when the chemical potential is positive. For these calculations, we set $\Delta_{SC}=0.01 eV$ and $M_z=0.02 eV$, related parameters of superconducting pairing attenuation consistent with those used for Bi$_2$Se$_3$. The corresponding BdG spectra and Wilson loops for these three topological superconducting states are shown in Fig.~\ref{InSb} c, d, Fig.~\ref{InSb} e, f and Fig.~\ref{InSb} g, h, respectively. It is noteworthy that the absolute value of the Chern number for all three states is 1, as there is only one $\Gamma$, X, and Y point in the first Brillouin zone.

\section{Conclusions}

In conclusion, we have developed a program that is integrated into WannierTools~\cite{wtgithub} to compute the superconducting BdG spectrum and superconducting topological invariants of 2D systems. We demonstrate its application in analyzing the topological superconducting properties of superconductor/candidate material/ferromagnetic insulator heterostructures. As a specific example, we show that SnTe can be utilized to construct chiral topological superconductors with a high Chern number. These results validate the program's functionality and rationale, positioning it as a valuable tool for first-principles calculations in the design and prediction of topological superconductors. Furthermore, this program provides theoretical support for the exploration of candidate materials for topological superconductors. Additionally, the computational framework can be extended to investigate other topological superconducting systems, such as magnetic topological insulator/superconductor heterostructures.

\section{Acknowledgments}

This work was supported by the National Key R\&D Program of China (Grant No. 2023YFA1607400), the National Natural Science Foundation of China (Grant No. 12274154, 12274436), the Science Center of the National Natural Science Foundation of China (Grant No. 12188101) and the Center for Materials Genome, China.


%

\end{document}